\documentclass[sigconf]{acmart}

\AtBeginDocument{%
  }

\usepackage[capitalise]{cleveref}
\copyrightyear{2026}
\acmYear{2026}
\setcopyright{cc}
\setcctype{by-nc-nd}
\acmConference[KDD '26]{Proceedings of the 32nd ACM SIGKDD Conference on Knowledge Discovery and Data Mining V.2}{August 09--13, 2026}{Jeju Island, Republic of Korea}
\acmBooktitle{Proceedings of the 32nd ACM SIGKDD Conference on Knowledge Discovery and Data Mining V.2 (KDD '26), August 09--13, 2026, Jeju Island, Republic of Korea}
\acmDOI{10.1145/3770855.3818373}
\acmISBN{979-8-4007-2259-2/2026/08}

\author{Difeng Ma}
\authornote{Work performed while the author was a research intern at StepFun.}
\authornote{Also with University of Chinese Academy of Sciences.}
\orcid{0009-0006-0364-1681}
\affiliation{%
  \institution{Computer Network Information Center, Chinese Academy of Sciences}
  \city{Beijing}
  \country{China}
}
\email{madifeng21@mails.ucas.ac.cn}

\author{Changhua Pei}
\authornote{Corresponding author, email:chpei@cnic.cn}
\authornote{Also with Hangzhou Institute for Advanced Study, University of Chinese Academy of Sciences.}
\affiliation{%
  \institution{Computer Network Information Center, Chinese Academy of Sciences}
  \city{Beijing}
  \country{China}
}
\email{chpei@cnic.cn}

\author{Yuanwei Lu}
\affiliation{%
  \institution{StepFun}
  \city{Shanghai}
  \country{China}
}
\email{luyuanwei@stepfun.com}

\author{Quan Zhou}
\affiliation{%
  \institution{Computer Network Information Center, Chinese Academy of Sciences}
  \city{Beijing}
  \country{China}
}
\orcid{0009-0000-3961-8877}
\email{zhouquan@cnic.cn}

\author{Zexin Wang}
\affiliation{%
  \institution{Computer Network Information Center, Chinese Academy of Sciences}
  \city{Beijing}
  \country{China}
}
\email{wangzexin@cnic.cn}

\author{Yibo Zhu}
\affiliation{%
  \institution{StepFun}
  \city{Shanghai}
  \country{China}
}
\email{zhuyibo@stepfun.com}

\author{Daxin Jiang}
\affiliation{%
  \institution{StepFun}
  \city{Shanghai}
  \country{China}
}
\email{djiang@stepfun.com}

\author{Dan Pei}
\affiliation{%
  \institution{Computer Science Department, Tsinghua University}
  \city{Beijing}
  \country{China}
}
\email{peidan@tsinghua.edu.cn}

\author{Jingjing Li}
\affiliation{%
  \institution{Computer Network Information Center, Chinese Academy of Sciences}
  \city{Beijing}
  \country{China}
}
\email{ljj@cnic.cn}

\author{Gaogang Xie}
\affiliation{%
  \institution{Computer Network Information Center, Chinese Academy of Sciences}
  \city{Beijing}
  \country{China}
}
\email{xie@cnic.cn}

\renewcommand{\shortauthors}{Difeng Ma et al.}
\usepackage{multirow}
\usepackage{booktabs}
\usepackage{xspace}
\usepackage[framemethod=TikZ]{mdframed}
\usepackage{appendix}
\usepackage{subcaption}
\usepackage{paralist}

\begin{document}

\title{Don't Predict, Prioritize: Rethinking GPU Reliability Assessment}



\newcommand{\HeaRank}{\textbf{HeaRank}\xspace}
\begin{abstract}
The reliability of Graphics Processing Units (GPUs) is a critical bottleneck for modern large-scale AI infrastructure, where a single node failure can disrupt synchronous training jobs and cause significant financial losses. While predictive maintenance is widely used in other hardware domains, we demonstrate that accurately predicting the exact timing of GPU failures is inherently difficult. Through an in-depth analysis of telemetry data from a production cluster, we find that major GPU failures, including Double Bit Errors (DBEs) and GPU Lost events, exhibit strong stochasticity and low signal-to-noise ratios in time-series telemetry, which makes conventional time-based prediction ineffective.

This insight motivates a paradigm shift: instead of attempting to predict the absolute timing of a failure, we propose a more robust approach focused on ranking nodes by their relative failure risk. We propose \HeaRank(Health Rank), a Learning-to-Rank (LTR) framework that leverages stable historical failure patterns to compute a global risk ranking of GPU nodes. Evaluated on a production-scale cluster with thousands of GPUs, \HeaRank achieves an AUC of 0.83, significantly outperforming both heuristic baselines and state-of-the-art ranking algorithms. In online deployment, HeaRank successfully captures 64\% of future failures within the top 5\% of ranked nodes, compared to only 21\% by the incumbent production system. These results suggest that relative risk ranking can serve as a robust alternative in environments where absolute failure prediction is inherently limited. Our work highlights the importance of risk-aware scheduling and proactive resource management in modern GPU clusters.
\end{abstract}




\maketitle

\section{Introduction}
Large-scale AI model training has become increasingly dependent on distributed GPU clusters, where thousands of GPUs work in coordination through synchronous training protocols. This tight coupling creates a critical vulnerability: Unexpected GPU failures can lead to significant service disruption, resource underutilization, and operational overhead~\citep{oles2024understanding,kokolis2025revisiting,dos2022experimental,deng2025minder}. Recent production evidence underscores this challenge: during Meta's LLaMA-3 training on 16,384 GPUs, over 400 job interruptions occurred, with nearly 60\% attributed to hardware failures~\citep{grattafiori2024llama}. Such widespread disruptions highlight that GPU failure mitigation has evolved from an operational concern to a fundamental requirement for scalable AI infrastructure~\citep{cui2025characterizing}.

To reduce the expenses associated with GPU failure, existing research attempts to forecast failure probabilities in advance of an actual breakdown~\citep{sun2019system,nikitin2022human}. In this paradigm, time-series telemetry metrics such as temperature, utilization, power consumption, and memory usage are continuously monitored, and machine learning models are trained to detect patterns indicative of impending failures. This methodology has proven remarkably effective across traditional domains such as manufacturing, hard disks, and even data center hard drives~\citep{xiao2018disk,li2016being,hamerly2001bayesian}.  For instance, hard disk drives demonstrate declining read/write speeds, increasing bad sector counts, and abnormal SMART metrics before failure~\citep{pinheiro2007failure, miller2023hard, tomer2021hard}. Similarly, optical transceivers show characteristic power degradation patterns, transmit power drops and receive sensitivity deteriorates, providing clear failure indicators~\citep{mou2025adaptive}. Memory modules manifest increasing ECC error rates and access latency spikes prior to permanent failures~\citep{cojocar2019exploiting,lv2020efficient,sridharan2015memory}. The fundamental reason for this success lies in the observability of hardware degradation: most hardware components exhibit clear, measurable precursor patterns before failure, allowing predictive algorithms to reliably detect impending failures.

Inspired by these successes, recent work has sought to apply similar techniques to GPU clusters~\citep{nie2018machine, gupta2025gpu, liu2023predicting}. However, unlike the well-established understanding of failure patterns in traditional domains, the applicability of predictive maintenance to GPUs remains largely uncharacterized. Specifically,  no prior work has provided direct evidence of consistent precursor patterns preceding GPU failures in production-scale clusters: the fundamental prerequisite that enables predictive maintenance in traditional domains. 

To address this gap, we conduct a systematic empirical investigation into the predictability of GPU failures utilizing thousands of real-world failure instances from our large-scale production environment. Our study focuses on the most critical GPU failure types: Double Bit  Errors (DBEs) and GPU Lost events, which represent uncorrectable memory corruption and host-communication loss, respectively~\citep{nvidiaXid}, and collectively account for 43.3\% of all service-impacting hardware incidents in our environment. We train five representative machine learning models (including XGBoost, CNN, LSTM, Transformer, and Mixture of Experts (MoE) ) based on telemetry observation windows preceding these failures. Across all models and configurations, failure prediction performance remains consistently poor and often approaches random guessing.

More importantly, our analysis reveals that this failure is not due to insufficient model capacity or feature engineering. Instead, it reflects fundamental properties of GPU operational data. Telemetry signals are heavily entangled with dynamic workloads, leading to strong non-stationarity and low signal-to-noise ratios. Failure precursors, when they exist, are temporally diffuse rather than localized, making them difficult to capture with fixed observation windows. Finally, the statistical distributions of telemetry metrics before failures are largely indistinguishable from those during normal operation. Together, these factors impose intrinsic limits on telemetry-based failure time prediction in real-world GPU clusters.

The inherent unpredictability of telemetry data compels us to look beyond transient metrics. Our analysis reveals a critical spatial insight: failures follow a Pareto distribution, concentrating on a small subset of "fragile" machines rather than being uniformly distributed. This suggests that historical reliability is a far more robust indicator than real-time volatility, motivating a paradigm shift from predicting \textbf{when} a failure will occur to prioritizing \textbf{which} nodes are most vulnerable. By reformulating the problem as risk ranking, we fundamentally improve the supervision structure. Unlike classification, which struggles with artificial temporal boundaries and label ambiguity, ranking leverages the relative ordering of nodes and accommodates longer observation horizons. This transforms sparse, stochastic failure events into dense, stable supervision signals, making risk prioritization a statistically feasible objective where precise temporal prediction fails.

Guided by this ranking-centric reformulation, we adopt Learning-to-Rank (LTR) models~\citep{liu2009learning, burges2005learning, burges2010ranknet} to leverage the inherent stability of historical failure records. Unlike volatile telemetry that requires complex temporal modeling, historical data directly encodes the relative health trends that LTR optimizes. To implement this, we introduce \textbf{\HeaRank}, a framework that aggregates cumulative failure signals (including counts, recovery patterns, and static metadata) to suppress stochastic fluctuations. Given that these historical features are highly discriminative, \HeaRank employs a straightforward Multi-Layer Perceptron (MLP) architecture. This design choice aligns with our core thesis that effective mitigation stems from principled problem formulation rather than architectural complexity. In online experiments, this robust approach achieves an AUC of 0.834 and captures 64\% of future failures within the top 5\% of ranked servers, demonstrating superior practicality over complex predictive baselines.

In summary, We make the following contributions:
\begin{compactitem}
\item We identify fundamental limitations in GPU failure prediction, revealing that time-series telemetry data is often dominated by noise, making accurate failure prediction highly unreliable.
\item We propose a paradigm shift from prediction to prioritization, casting proactive GPU failure mitigation as a LTR task focused on node prioritization rather than precise prediction. We implement HeaRank, a lightweight MLP model that leverages stable historical features and demonstrates its superior practical outcomes.

\item We validate our approach on a production-scale cluster. It outperforms multiple baselines,
including failure count heuristics, an existing health score system, and LightGBM Ranker, achieving an AUC of \textbf{0.834}. In deployment, HeaRank captures \textbf{64\%} of future failures within the top 5\% of ranked GPU servers, versus \textbf{21\%} for the existing system.
\end{compactitem}

\section{Related Work}
\noindent\textbf{GPU Failure Prediction and Reliability.}
Prior work has studied GPU reliability and failure prediction in production and HPC environments. Nie et al.~\citep{nie2018machine} investigate machine learning models for predicting GPU errors in large-scale HPC systems, highlighting the challenges of data imbalance and hardware heterogeneity. Liu et al.~\citep{liu2023predicting} propose deep learning based methods to predict GPU failures under deep learning workloads, demonstrating that certain failure types can be predicted with high precision under specific settings. Complementary measurement studies characterize GPU error modes and root causes, including memory corruption and unrecoverable errors~\citep{dos2022experimental,oles2024understanding,cui2025characterizing}. At the system level, recent work emphasizes that reliability has become a first-order concern for large-scale model training, where a single device failure can disrupt synchronized jobs and waste substantial compute resources~\citep{kokolis2025revisiting,deng2025minder}.

\noindent\textbf{Learning-to-Rank (LTR) and Ranking Baselines.}
Learning-to-Rank is a mature machine learning paradigm that learns an ordering over items, and it is widely used in information retrieval and recommendation systems~\citep{liu2009learning}. Classical methods include pairwise approaches such as RankNet and LambdaRank, and tree-based ranking models such as LambdaMART, which optimize ranking metrics directly~\citep{burges2005learning,burges2010ranknet}. In this work, we leverage LTR as a more operationally aligned formulation for risk assessment, where the goal is to prioritize machines by relative risk rather than to predict an exact failure time.

Operational practice often relies on simple heuristics derived from historical incidents, for example ranking machines by total failure count or by weighted counts that assign higher weights to severe failure types. Many production systems also maintain a composite health score based on domain knowledge and monitoring rules. In our evaluation, we include these three representative baselines, unweighted failure count ranking, weighted failure count ranking, and an existing health score system, as they reflect common, easy-to-deploy strategies in real clusters. We also compare against LightGBM Ranker~\citep{lightgbm2017highly}, a strong learning-based ranking baseline. Beyond these, simple linear scorers (e.g., logistic regression and linear SVM) are often used as quick risk-scoring baselines, and survival analysis methods such as Cox proportional hazards~\citep{cox1972regression} and Random Survival Forest~\citep{ishwaran2008random} provide an alternative statistical framework that directly models time-to-event and naturally yields a risk ranking. We include representative members of these families to ensure a broader comparison.

\section{Predictability Analysis}
\subsection{Model Prediction Experiments}\label{sec:model}

\noindent\textit{Problem formulation.} We cast GPU failure prediction as a time-series classification task. For each GPU, we extract a fixed-length observation window $\mathbf{X}_{t-L:t}$ from telemetry streams and predict whether a failure will occur within the subsequent horizon $\Delta t$. Each window is assigned a binary label based on whether a failure occurs within $\Delta t$, and models are trained to map $\mathbf{X}_{t-L:t}$ to this label.

\noindent\textit{Results.} We evaluated five representative models, XGBoost, CNN, LSTM, Transformer, and Mixture of Experts (MoE), under multiple horizons and window sizes. Across all settings, prediction performance remains consistently poor and often approaches random guessing. As shown in Table~\ref{tab:model_performance}, even the top-performing MoE model achieves a maximum F1-score of only 0.4837. These results suggest that the limitation is not model capacity, but the intrinsic properties of production telemetry. Detailed dataset construction and preprocessing are provided in Appendix.

\begin{mdframed}[
    leftmargin  = 0pt,    
    rightmargin = 0pt,    
    innerleftmargin = 8pt,
    innerrightmargin= 8pt,
    innertopmargin = 6pt, 
    innerbottommargin = 6pt,
    skipabove = 8pt,      
    skipbelow = 8pt,      
    linewidth = 0.8pt     
]
\textbf{\textit{Insight 1: GPU failures are both sparse and stochastic. 
This makes failure prediction fundamentally unreliable under precursor-based approaches.}}
\end{mdframed}
This insight is corroborated by the intrinsic characteristics of GPU telemetry, where signal patterns are inextricably coupled with workload dynamics. AI tasks, including training, inference, and fine-tuning, exhibit inherently volatile and periodic behavior. These workload characteristics induce sharp fluctuations in metrics such as power and utilization, even during healthy operation. Furthermore, the combination of workload heterogeneity and opaque scheduling means that metrics from healthy machines often follow vastly different statistical distributions, complicating the establishment of a unified baseline for anomaly detection.
\begin{figure}[t]
    \centering
    \includegraphics[width=0.95\linewidth]{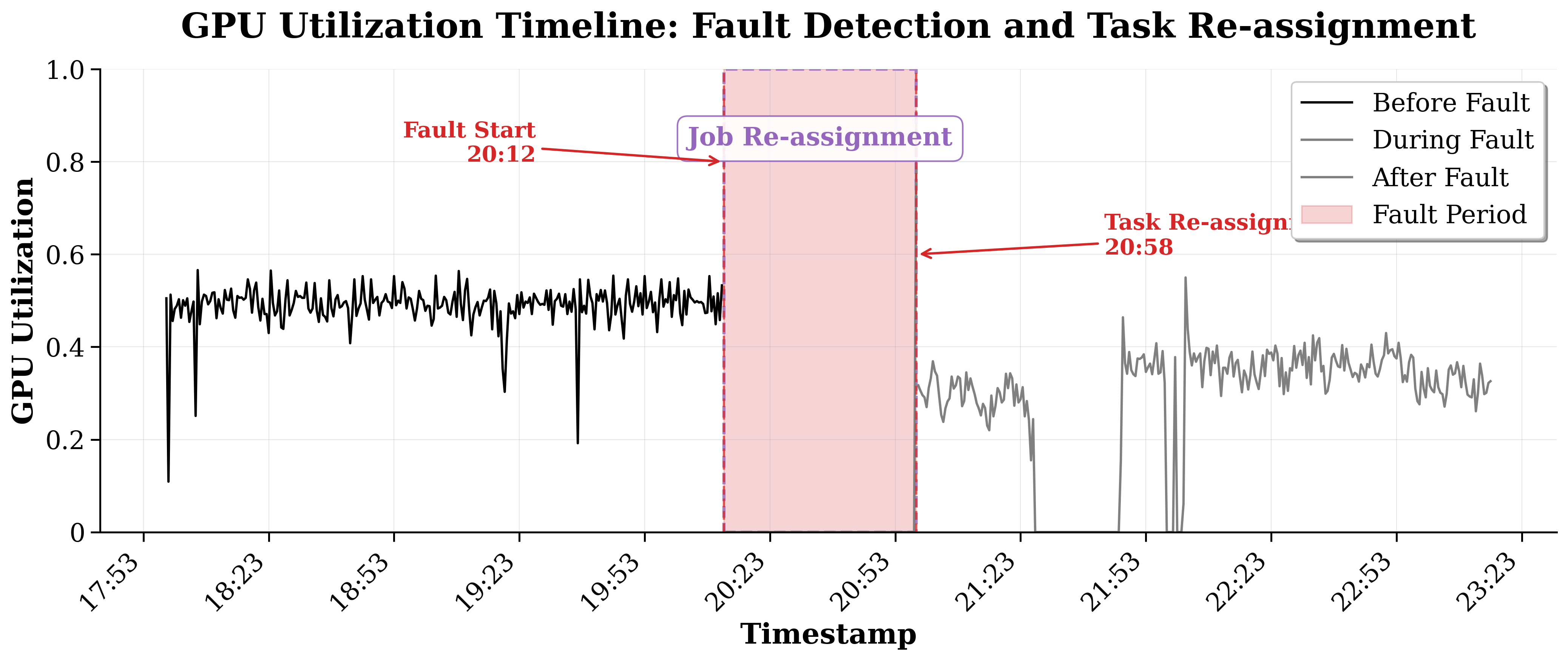}
    \vspace{-0.3cm} 
    \caption{Workload redistribution breaks temporal continuity. The post-failure context shift (red zone) introduces noise that confounds predictive modeling.}
    \label{fig:gpu_timeline}
    \vspace{-0.4cm} 
\end{figure}

\begin{figure}[htbp]
    \centering
    \includegraphics[width=0.9\linewidth]{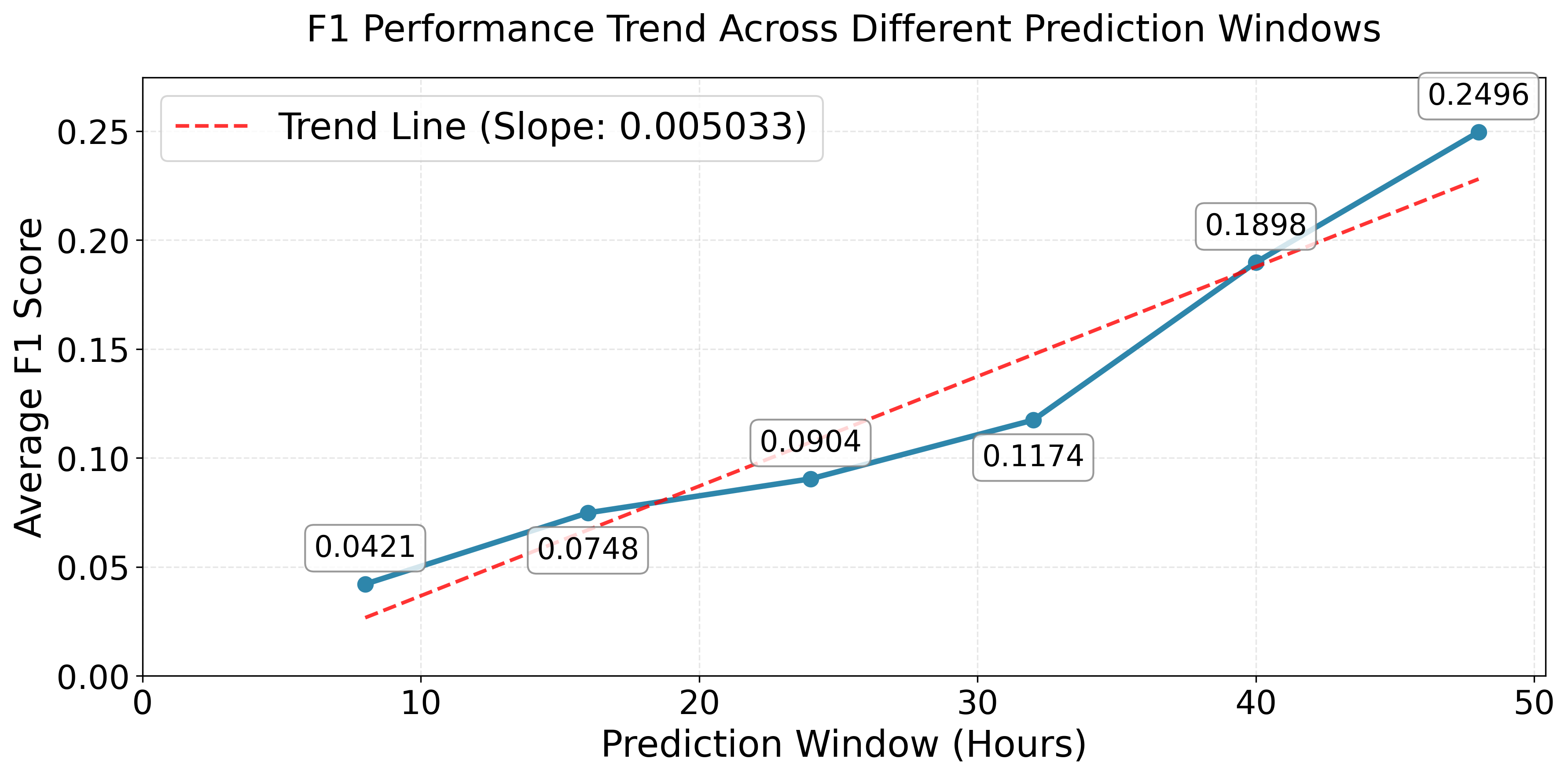}
    \vspace{-0.3cm}
    \caption{Rising F1 scores with longer history indicate diffusive precursors. The lack of a consistent lead time undermines fixed-window prediction.}
    \label{fig:f1trend}
    \vspace{-0.5cm}
\end{figure}
This challenge is compounded by post-failure operational shifts (Figure~\ref{fig:gpu_timeline}). Following critical events like DBEs or GPU Lost, schedulers typically reassign the affected node to different workload pools. This \textit{distributional shift} prevents the accumulation of failure samples under consistent operational conditions, making it nearly impossible for models to learn stable precursor patterns. Combined with the extreme sparsity of critical failures, these factors render signal extraction exceptionally difficult.

Beyond signal noise, valid precursor patterns lack temporal localization. As shown in Figure~\ref{fig:f1trend}, extending the observation window for the MoE model from 8 to 48 hours leads to a steady increase in F1-score. This trend suggests that failure precursors are \textit{temporally ambiguous} and do not manifest at a consistent lead time. This ambiguity invalidates standard sliding-window supervision, which assumes a fixed temporal relationship between past observations and future failures. In practice, fixed-length windows frequently miss the actual precursor onset, leading to our second key insight:

\begin{mdframed}[
    leftmargin = 0pt,
    rightmargin = 0pt,
    innerleftmargin = 8pt,
    innerrightmargin= 8pt,
    innertopmargin = 6pt,
    innerbottommargin = 6pt,
    skipabove = 8pt,
    skipbelow = 8pt,
    linewidth = 0.8pt
]
\textbf{\textit{Insight 2: Temporal precursors are diffusive and difficult to localize. 
The lack of a consistent lead time undermines conventional time-windowed prediction, as fixed observation windows fail to capture the complete evolution of failure risks.}}
\end{mdframed}

This consistent underperformance, regardless of architecture or horizon, confirms that the bottleneck lies not in algorithmic capacity, but in the inherent unpredictability of the telemetry data itself.

\subsection{Data Analysis}\label{sec:data_analysis}

To understand why predictive models fail despite diverse architectures and window settings, we analyze the intrinsic properties of GPU telemetry data. Rather than focusing on model behavior, we examine whether telemetry signals themselves exhibit (i) consistent temporal trends, (ii) sufficient signal quality, and (iii) distinguishable distributions prior to failures. Using correlation analysis, signal-to-noise measurements, and distributional comparisons, we show that these prerequisites are largely violated in production environments. Together, these results indicate that telemetry data alone lacks the structural properties required for reliable early failure prediction.

\subsubsection{Failure Pattern Consistency Analysis}
We first examine whether GPU telemetry metrics exhibit consistent temporal trends prior to failure events. To this end, we measure the monotonic correlation between metric values and time-to-failure using Kendall's rank correlation coefficient ~\citep{abdi2007kendall}.

\begin{figure}[tbp]
\centering
\includegraphics[width = 0.8\linewidth]{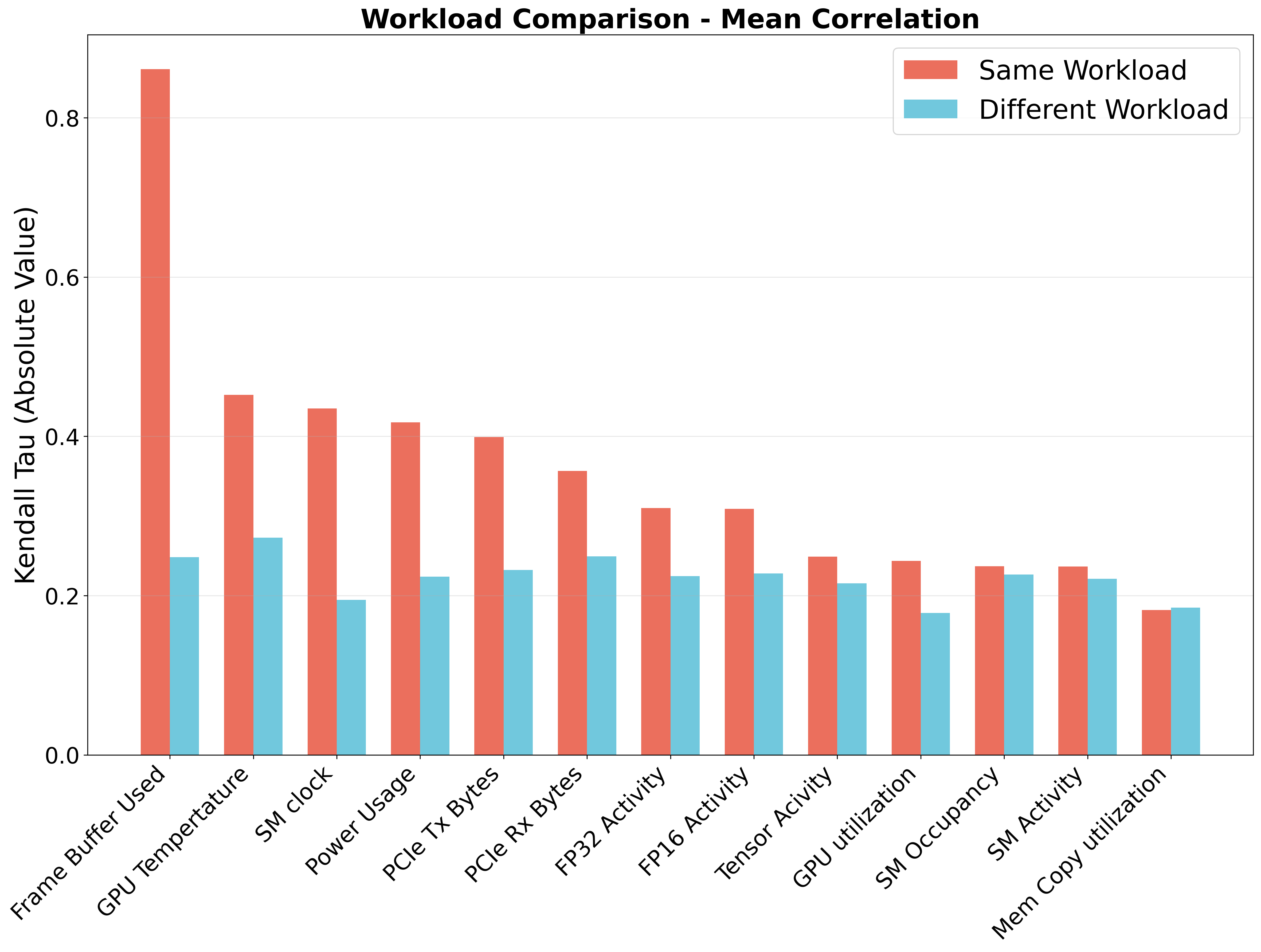}
\vspace{-0.3cm} 
\caption{\enspace Kendall correlation coefficient analysis results for different GPU timing metrics, two colors indicate data sources under different settings.}
\vspace{-0.4cm} 
\label{fig:kendall comparison}
\end{figure}

\begin{figure}[tbp]
\centering
\includegraphics[width = \linewidth]{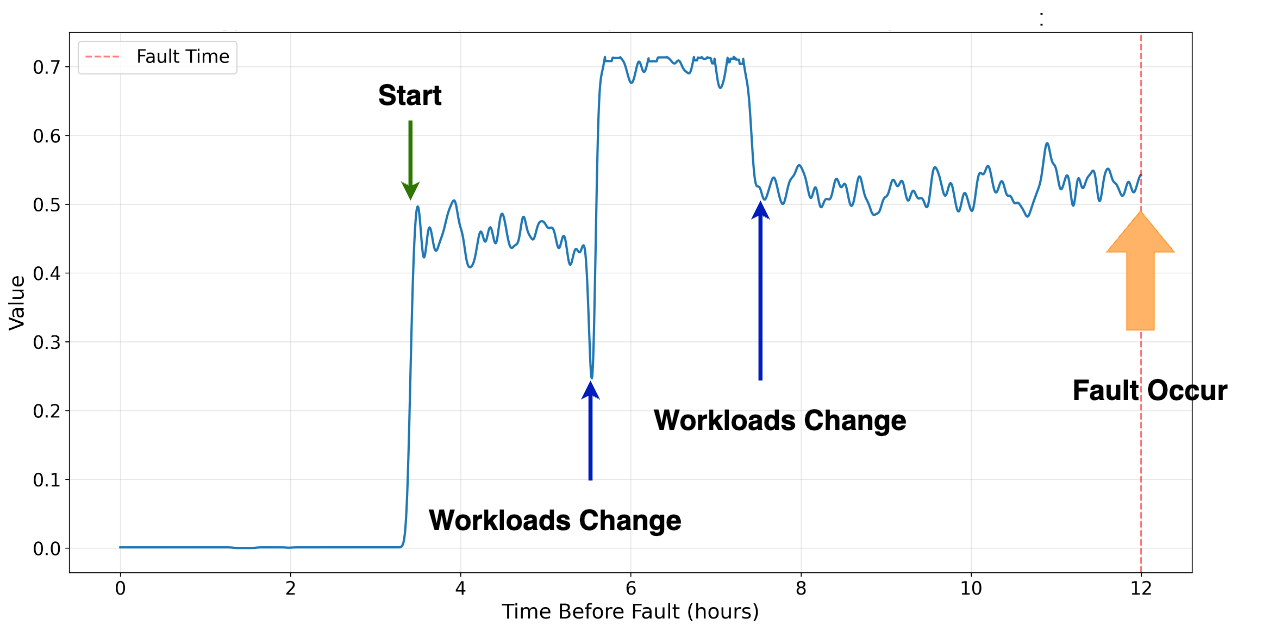}
\caption{\enspace Changes in timing curves when switching between different workloads }
\label{fig:workload_shift}
\vspace{-0.5cm}
\end{figure}

Figure~\ref{fig:kendall comparison} reveals a striking difference: measurable temporal correlations emerge only under conditions where workloads remain stable throughout the observation window. Once workload changes occur, correlation values across all metrics rapidly degrade toward zero, indicating that apparent precursor trends are highly fragile and workload-dependent rather than failure-intrinsic. This stark contrast identifies workload variability as the primary factor disrupting patterns. The mechanism behind this disruption is illustrated in Figure~\ref{fig:workload_shift}. When workloads change, the resulting trend shifts introduce non-stationarity that makes it nearly impossible for models to learn consistent precursor patterns~\citep{oles2024understanding}. This effect is particularly pronounced for memory-intensive metrics. Under stable workloads, Frame Buffer Used achieves a Kendall $\tau$ coefficient of 0.861 for DBE failures, reflecting the strong correlation between sustained memory stress and bit-level degradation~\citep{oles2024understanding}. However, such ideal conditions represent only a small fraction of real-world scenarios, limiting the generalizability of this finding. 

This behavior reveals a fundamental limitation of temporal failure prediction in production environments. In real-world clusters, workload migration and scheduling dynamics routinely introduce non-stationarity, breaking the temporal continuity required for models to learn reliable precursors. As a result, even when certain metrics display strong correlations under controlled settings, such patterns fail to generalize to operational scenarios, undermining the practicality of trend-based failure prediction.

\subsubsection{Quantifying Data Randomness.}
Beyond workload variability, we quantify data quality using Signal-to-Noise Ratio (SNR), where "signal" represents the low-frequency trend extracted via Gaussian filtering, and "noise" captures high-frequency fluctuations.

\begin{figure}[tbp]
\centering
\includegraphics[width = 0.8\linewidth]{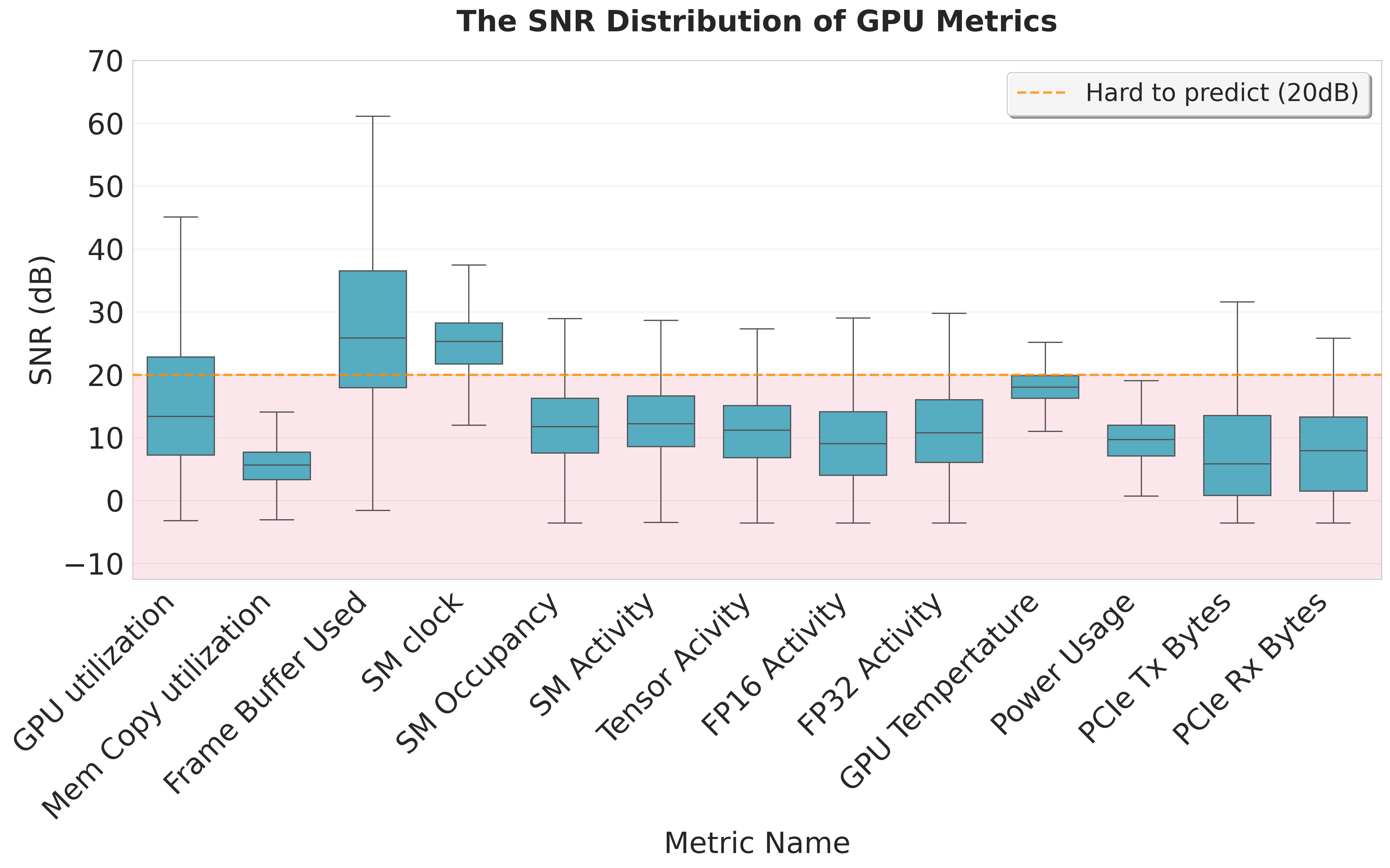}
\caption{\enspace Boxplot of SNR signal-to-noise ratio analysis for different GPU metrics.}
\vspace{-0.6cm}
\label{fig:snr}
\end{figure}

 Figure~\ref{fig:snr} reveals a clear dichotomy in data quality across GPU metrics. Hardware-controlled metrics (Frame Buffer Used, SM Clock, Temperature) exhibit high SNR values, reflecting their stable, regulation-driven behavior. Conversely, metrics directly tied to dynamic application activity (PCIe Tx/Rx, Tensor Activity) show extremely low SNR, indicating they are dominated by workload-induced noise rather than meaningful temporal patterns.

This analysis exposes a second fundamental challenge: even when workload variability is controlled, most GPU metrics lack the signal quality necessary for reliable early warning systems. The pervasive noise in telemetry data means that even sophisticated deep learning models struggle to distinguish between meaningful failure precursors and random fluctuations. 

\subsubsection{Distribution Analysis}
A necessary precondition for telemetry-based failure prediction is that pre-failure behavior is statistically distinguishable from normal operation~\citep{chandola2009anomaly,liu2013changepoint}. To test this, we compare the empirical distributions of key metrics collected during the 24 hours preceding failures with those from normal periods on non-failing GPUs. Our goal is to determine whether pre-failure samples exhibit consistent distributional shifts that would make failure and non-failure states separable in practice.

\begin{figure*}[htbp]
\centering
\includegraphics[width = 0.8\linewidth]{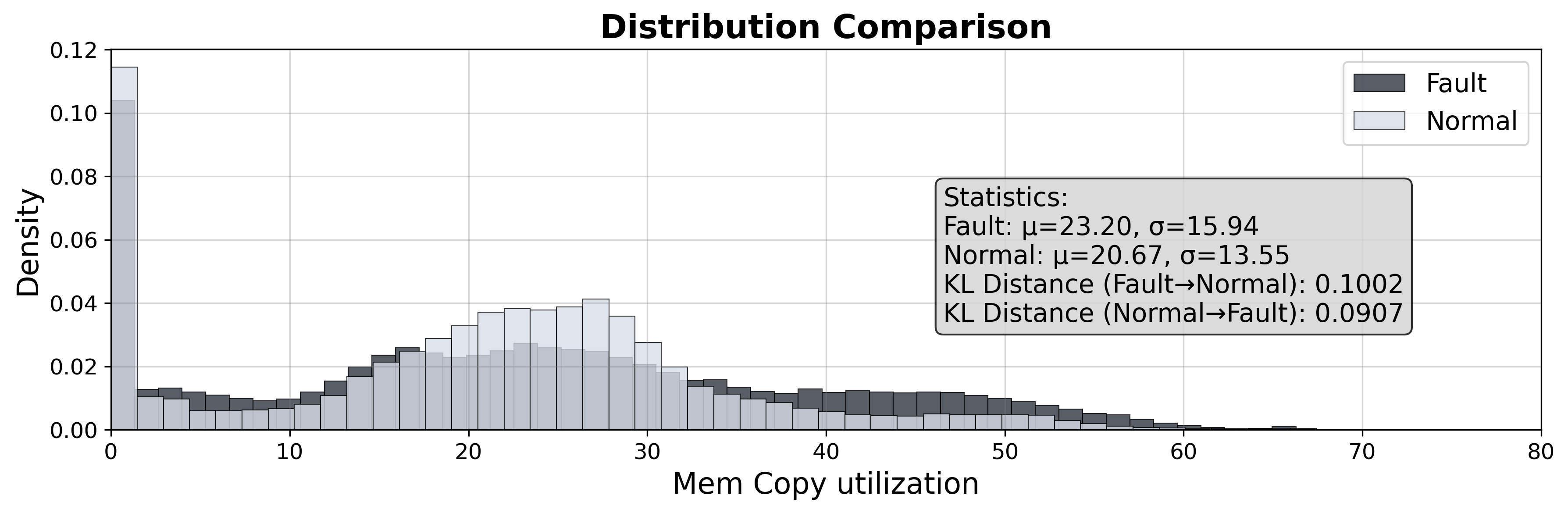}
\caption{\enspace Statistical distribution comparison between pre-failure periods (24 hours before failure) and normal operation for GPU telemetry metrics. The substantial overlap indicates that failure and normal states are statistically indistinguishable, challenging traditional anomaly detection approaches.}
\label{fig:Distribution comparison}
\vspace{-0.5cm}
\end{figure*}

Figure~\ref{fig:Distribution comparison} shows that such shifts are largely absent. Across the examined metrics, pre-failure and normal samples exhibit highly overlapping distributions, with nearly identical central tendencies and variances. This strong overlap indicates that, at least at the level of marginal statistics, pre-failure telemetry is not reliably separable from normal behavior, limiting the effectiveness of distribution-based predictors.

This result suggests a third fundamental challenge: pre-failure telemetry is statistically indistinguishable from normal operation, making reliable failure prediction via classification infeasible.

\subsection{The Power of Historical Failure Patterns}
The above observations compel us to rethink GPU time-series failure prediction from a new perspective, rather than treating it as a conventional precursor-mining task. At first glance, it is tempting to use historical failure information (e.g., counts, recency, and inter-failure intervals) as predictors for time-local failure prediction. However, these signals are inherently \textit{longitudinal} across GPU instances: they are sparse in time, differ substantially across GPUs, and change slowly within a given GPU. In contrast, standard telemetry metrics are \textit{cross-sectional} within each GPU: they can vary sharply across time windows for the same GPU due to workload dynamics.

This granularity mismatch creates two practical limitations. First, using failure-history features alone to predict \textit{when} a failure will occur is inherently imprecise because the signals are temporally sparse and do not provide fine-grained localization. Second, when combined with window-level telemetry under the standard sliding-window formulation, failure-history features often remain nearly identical across many windows of the same GPU, while window labels differ (positive vs. negative) depending on proximity to failures. This produces contradictory supervision at the sample level, preventing models from extracting meaningful information from historical failure features.

Importantly, aggregated historical features also serve as a form of \textit{denoising}: by summarizing long-term incident records (e.g., counts, recency, and inter-failure intervals), they act as a low-pass filter that suppresses short-term workload-induced fluctuations present in raw telemetry. This makes the learning signal more stable and better aligned with persistent host-level fragility.

\begin{figure}[htbp]
\centering
\includegraphics[width = 0.8\linewidth]{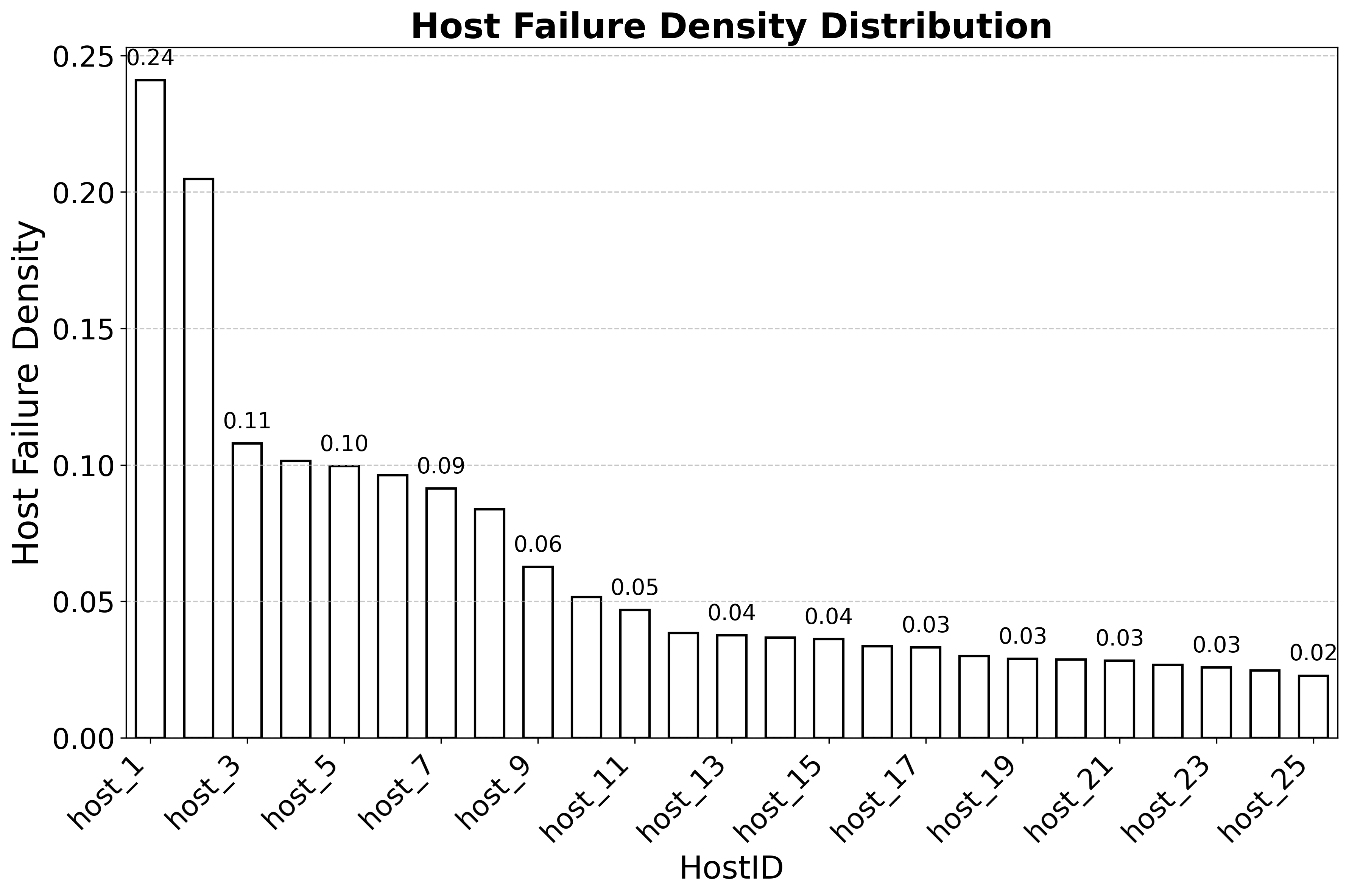}
\caption{\enspace Failure concentration follows a Pareto distribution: a small subset of "fragile" hosts accounts for the majority of critical incidents, revealing stable risk patterns that persist over time.}
\label{fig:host density}
\vspace{-0.5cm}
\end{figure}
\begin{figure}[tbp]
\centering
\includegraphics[width = 0.8\linewidth]{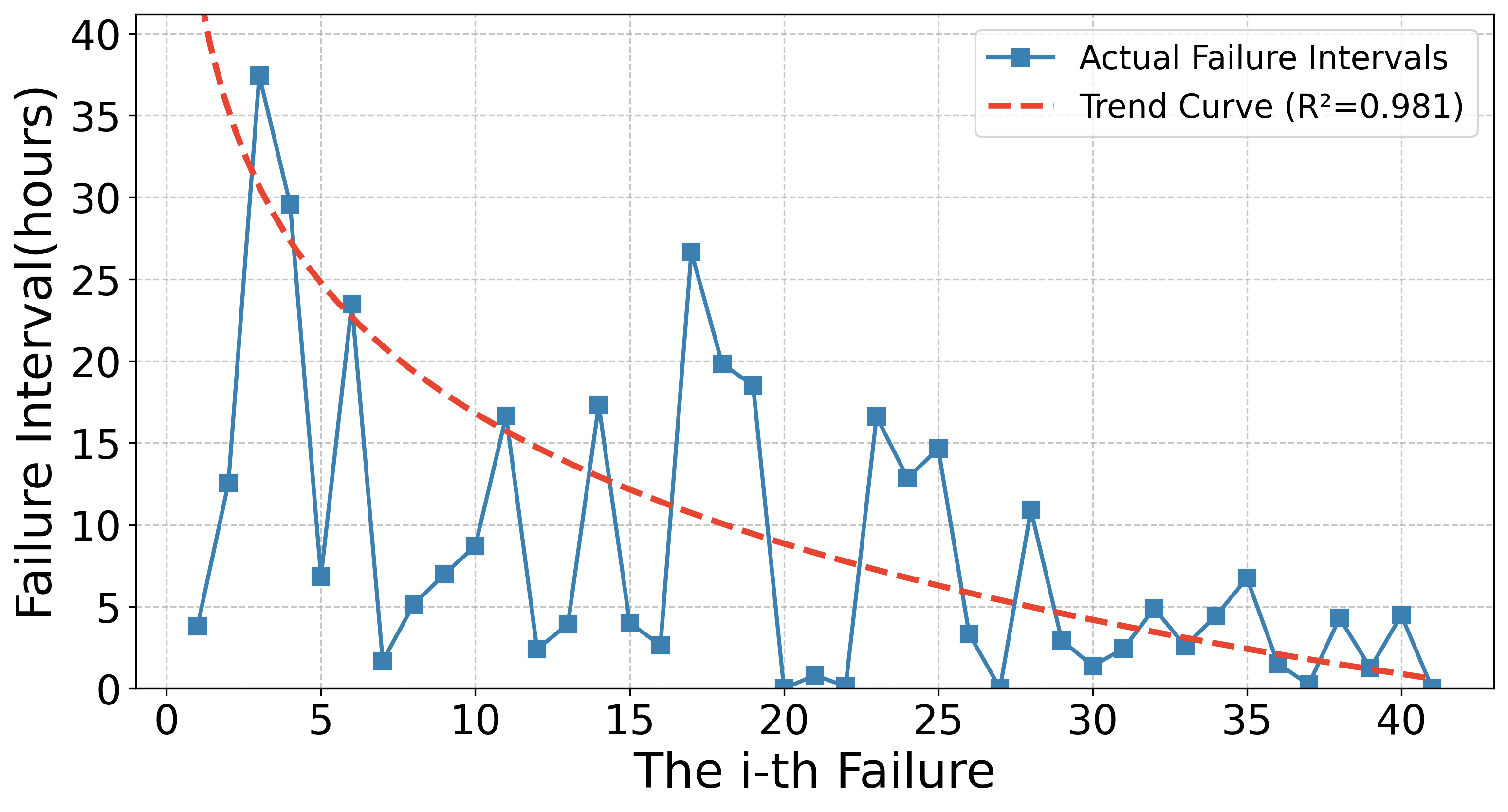}
\vspace{-0.4cm}
\caption{\enspace Distribution of time intervals between consecutive failures for high-risk machines during a peak failure month. The prevalence of short intervals confirms that past failures are strong predictors of future failures.}
\label{fig:failure intervals}
\vspace{-0.5cm}
\end{figure}

While micro-level telemetry proved unreliable, macro-level analysis of cluster failure history revealed a remarkably stable and actionable pattern. Failures exhibit a clear Pareto distribution~\citep{arnold2014pareto}: over 30\% of all critical DBE and GPU Lost incidents originate from less than 10\% of cluster hosts (Figure~\ref{fig:host density}). A $\chi^2$ goodness-of-fit test against a Poisson null rejects randomness with $p \ll 10^{-10}$, and this concentration is temporally stable: the top 10\% of hosts consistently cover roughly 24--33\% of failures across quarters and rolling 90-day windows. This concentration is not random but reflects underlying hardware fragility and operational vulnerabilities. Figure~\ref{fig:failure intervals} further demonstrates failure recurrence patterns within high-risk hosts. The clustering of short inter-failure intervals indicates that once a machine experiences a critical failure, subsequent failures tend to follow rapidly. This phenomenon suggests that either persistent latent hardware issues or insufficiently resolved root causes during recovery.

This analysis reveals a crucial insight that transforms our approach:
\begin{mdframed}[
    leftmargin  = 0pt,    
    rightmargin = 0pt,    
    innerleftmargin = 8pt,
    innerrightmargin= 8pt,
    innertopmargin = 6pt, 
    innerbottommargin = 6pt,
    skipabove = 8pt,      
    skipbelow = 8pt,      
    linewidth = 0.8pt     
]
\textit{\textbf{Insight 3:}} \textbf{\textit{while historical failure information is indeed predictive of machine risk, its utility lies at the host level, not at the fine-grained sample level. }}
\end{mdframed}
While reliable short-term precursors are scarce in time-series telemetry, a node's long-term health can be robustly assessed using its failure history. Servers that have failed previously are significantly more likely to fail again, consistent with the well-documented "lemon nodes" phenomenon~\citep{cui2025characterizing}.

This insight motivates a fundamental paradigm shift from time-local prediction to risk ranking. This change is driven not only by the fact that historical failure information is naturally suited to host-level ranking, but also because ranking better matches operational needs: knowing exactly when a node will fail is often less useful than proactively reducing the impact of failures. Rather than discarding high-risk machines, risk ranking enables priority-aware scheduling. High-priority jobs can be placed on lower-risk GPUs, while higher-risk nodes are steered toward lightweight or interruption-tolerant workloads, maximizing overall utilization. Meanwhile, host risk scores can be updated continuously as new failures occur, allowing the scheduler to dynamically adapt resource allocation and maintenance actions~\citep{stokely2025shaved,chiang2014profit,venkateswaran2017uptime}. 

\section{HeaRank Model}
\subsection{Problem Statement}
We formulate GPU failure risk assessment as a LTR problem. At each query time $q$ (e.g., daily evaluation), we aim to rank all $n$ machines in the cluster by their likelihood of failure within a prediction window $\Delta t$ (e.g., 7 days).

For each machine $i$, we define:
\begin{compactitem}
    \item feature vector $\mathbf{x}_i \in \mathbb{R}^m$, summarizing historical failure patterns and system characteristics;
    \item binary label $y_i \in \{0,1\}$, where $y_i = 1$ indicates that machine $i$ experiences a failure within the next $\Delta t$.
\end{compactitem}

Our goal is to learn a scoring function $S(\mathbf{x}_i)=f(\mathbf{x}_i;\theta)$ such that machines with higher scores are ranked as higher risk (i.e., more likely to fail within $\Delta t$).

\subsection{Model Overview}
Building on the insight that historical failure information is a stable and predictive proxy for machine health, we design a proactive risk-ranking system named HeaRank. Rather than predicting exact failure timestamps from volatile telemetry, HeaRank summarizes each host's long-term incident history and context into a global risk score that can be directly consumed by schedulers and maintenance workflows.

At the model core, HeaRank employs a \textit{Non-linear Risk Interaction Encoder}, a compact multi-layer perceptron (hidden sizes 512, 512, 256 with ReLU and batch normalization) that maps engineered host features into a latent risk representation. We adopt an MLP (instead of a linear scorer) because failure risk is driven by \textit{cross-view, non-linear interactions} among heterogeneous signals (static host profile, long-term failure history, and short-term recency). The MLP acts as a contextual denoiser: it learns to distinguish between 'transient instabilities' and 'persistent degradation'. While high recency generally implies high risk, the Encoder modulates this signal using static host embeddings, effectively suppressing false alarms from robust nodes that experienced incidental errors.

Given the encoder output, HeaRank predicts a scalar risk probability via a probability calibration head, and we train it with a pointwise sigmoid (cross-entropy) objective. Importantly, this choice is dictated by two production requirements. In operations we need not only an ordering, but also an \textit{absolute} risk probability to support globally consistent thresholding policies (e.g., "investigate the top 5\% highest-risk hosts"), alerting, and fleet-level capacity planning. Pointwise sigmoid training directly learns a calibrated probability interpretation for $\Pr(y=1\mid \mathbf{x})$, enabling stable decision thresholds across time. The pointwise formulation yields $O(N)$ scoring, allowing fast daily (or more frequent) re-ranking at fleet scale. Empirically, we also find that swapping the pointwise objective for listwise (ListNet) or pairwise (LambdaRank-style) supervision does not consistently improve ranking quality on our data; the full comparison is deferred to Appendix~\ref{appendix:ltr_objective}.

Training data is constructed from all recorded failures between January 2024 and December 2025. Every three days, we generate a query for each monitored machine; queries are labeled positive if a failure occurs within the next seven days and negative otherwise.

Formally, for a machine $i$ with feature vector $x_i$ and binary label $y_i \in \{0,1\}$, the loss is computed as:
\begin{align}
    \hat{y}_i &= \sigma(\mathbf{w}^\top \phi(\mathbf{x}_i) + b) \\
    L &= -\frac{1}{N} \sum_{i=1}^{N} \left[ y_i \log \hat{y}_i + (1 - y_i) \log (1 - \hat{y}_i) \right]
\end{align}
where $\phi(\cdot)$ is the Non-linear Risk Interaction Encoder, $\mathbf{w}, b$ are learnable parameters, and $\hat{y}_i$ denotes the predicted probability that host i will fail in $\Delta t$. $\sigma(\cdot)$ is the sigmoid activation function that converts the raw score into a probability.

\section{Evaluation}
This section aims to rigorously evaluate the practical effectiveness of our proposed ranking-based method, HeaRank, and provide answers to four key research questions:
\begin{compactitem}
    \item \textbf{RQ1}: How does HeaRank perform compared to existing heuristics and learning-based baselines?
	\item \textbf{RQ2}: Which features contribute most significantly to the effectiveness of machine risk ranking?
	\item \textbf{RQ3}: How sensitive is the model performance to the choice of supervision window used in training?
	 \item \textbf{RQ4}: Does augmenting historical features with aggregated telemetry further improve ranking quality? 
\end{compactitem}

\subsection{Baselines and evaluation metrics}

\noindent\textbf{Evaluation Metrics.}
We evaluate HeaRank using two complementary metrics: \textbf{AUC} and \textbf{NDCG@K}. AUC measures the model's overall discriminative power in a threshold-free manner (i.e., the probability that a randomly sampled failing host is assigned a higher score than a healthy one), while NDCG@K evaluates top-$K$ ranking quality with position-based discounting, emphasizing correctness at the very top of the list.

We choose NDCG@K as the primary metric because operations can only act on a small risk budget (e.g., investigate the top few percent hosts), and we report AUC to complement it with a global view of separability across the fleet.

\noindent\textbf{Baselines.} To evaluate HeaRank's effectiveness, we compare against two categories of baselines: operational heuristics reflecting standard practices, and a strong learning-based ranker.

\textbf{Operational Heuristics:} We evaluate three rule-based methods: (1) \textit{Unweighted Failure Count Ranking} that ranks hosts by total historical failures, (2) \textit{Weighted Failure Count Ranking} that assigns fixed weights to different failure types, and (3) the \textit{Existing Health Score System} currently deployed in production. While these methods provide operational simplicity, they inherently lack adaptability to evolving failure patterns and complex data relationships.

\textbf{Learning-based Baselines:} We employ LightGBM Ranker~\citep{lightgbm2017highly} as our primary machine learning baseline, a state-of-the-art gradient boosting framework for ranking tasks. To further address the concern that simple models may already suffice, we also include Logistic Regression and Linear SVM trained on the same host-level features. Finally, since survival analysis is a natural statistical framework for time-to-event ranking, we add Cox proportional hazards (CoxPH)~\citep{cox1972regression} and Random Survival Forest (RSF)~\citep{ishwaran2008random} as additional baselines.

\subsection{RQ1: Evaluation Results and Analysis}

Table~\ref{tab:model_comparison} highlights HeaRank's consistent advantage over all baselines. Simple rule-based heuristics and the Health Score baseline perform poorly, confirming the limited value of experience-driven rules for risk assessment. LightGBM Ranker, representing the state-of-the-art machine learning baseline, achieves reasonable AUC but falls short in top-ranked precision, which is 
critical for operations. 

HeaRank achieves both higher discriminative power (AUC 0.834 vs. 0.795) and substantially better high-precision ranking (NDCG@5 of 0.427 vs. 0.309, a 38\% relative gain). These improvements hold across all cutoffs, demonstrating robustness under different operational priorities. Among the added baselines, Logistic Regression is clearly weaker, while Linear SVM and Random Survival Forest are surprisingly competitive on NDCG, confirming that once the problem is reframed as host-level ranking, even simple scorers become viable. Still, HeaRank remains the most consistent across AUC and all top-$K$ cutoffs. The gains arise from
leveraging stable historical features and capturing feature interactions that traditional methods overlook.

\begin{table*}[htbp]
\centering
\small
\caption{Performance comparison of HeaRank with different baseline models. The NDCG values represent the mean and variance under different queries in the test dataset.}
\label{tab:model_comparison}
\begin{tabular}{lcccc}
\toprule
\textbf{Method} & \textbf{AUC} & \textbf{NDCG@5} & \textbf{NDCG@10} & \textbf{NDCG@20} \\
\midrule
Failure Count (Unweighted) & 0.6060  & 0.1383 $\pm$ 0.2187 & 0.1339 $\pm$ 0.1868 & 0.1331 $\pm$ 0.1752 \\
Failure Count (Weighted) & 0.6524  & 0.1522 $\pm$ 0.2364 & 0.1168 $\pm$ 0.1953 & 0.1687 $\pm$ 0.1835 \\
Health Score (Reversed) & 0.5820  & 0.0595 $\pm$ 0.1302 & 0.0629 $\pm$ 0.1242 & 0.0692 $\pm$ 0.1210 \\
\midrule
Logistic Regression & 0.6343 & 0.2901 $\pm$ 0.2729 & 0.2508 $\pm$ 0.2211 & 0.2886 $\pm$ 0.1374 \\
Linear SVM & 0.7860 & 0.4164 $\pm$ 0.2470 & 0.4435 $\pm$ 0.1894 & 0.4056 $\pm$ 0.1505 \\
CoxPH & 0.6948 & 0.1131 $\pm$ 0.1517 & 0.1207 $\pm$ 0.1057 & 0.0488 $\pm$ 0.0643 \\
Random Survival Forest & 0.8075 & 0.4503 $\pm$ 0.3439 & 0.3891 $\pm$ 0.2900  & 0.2233 $\pm$ 0.2407 \\
\midrule
LightGBM Ranker & 0.7947 & 0.3088 $\pm$ 0.3700 & 0.2902 $\pm$ 0.3200 & 0.2707 $\pm$ 0.3100 \\
\textbf{\HeaRank (Ours)} & \textbf{0.8340} & \textbf{0.4267 $\pm$ 0.3600} & \textbf{0.4225 $\pm$ 0.3000} & \textbf{0.3743 $\pm$ 0.2300} \\
\bottomrule
\end{tabular}
\end{table*}

\subsection{RQ2: Ablation Study}

To assess which information sources drive ranking quality, we ablate major feature categories one at a time
while keeping all other settings fixed. We group features into four coarse types: (i) \textit{failure
composition} (what kinds of faults a host tends to see), (ii) \textit{recurrence/recency statistics} (how often
and how recently faults recur), (iii) \textit{calendar-time context} (weak periodic patterns such as weekly
cycles), and (iv) \textit{host metadata} (persistent host-/hardware-specific identifiers).

\begin{figure}[htbp]
\centering
\includegraphics[width=\linewidth]{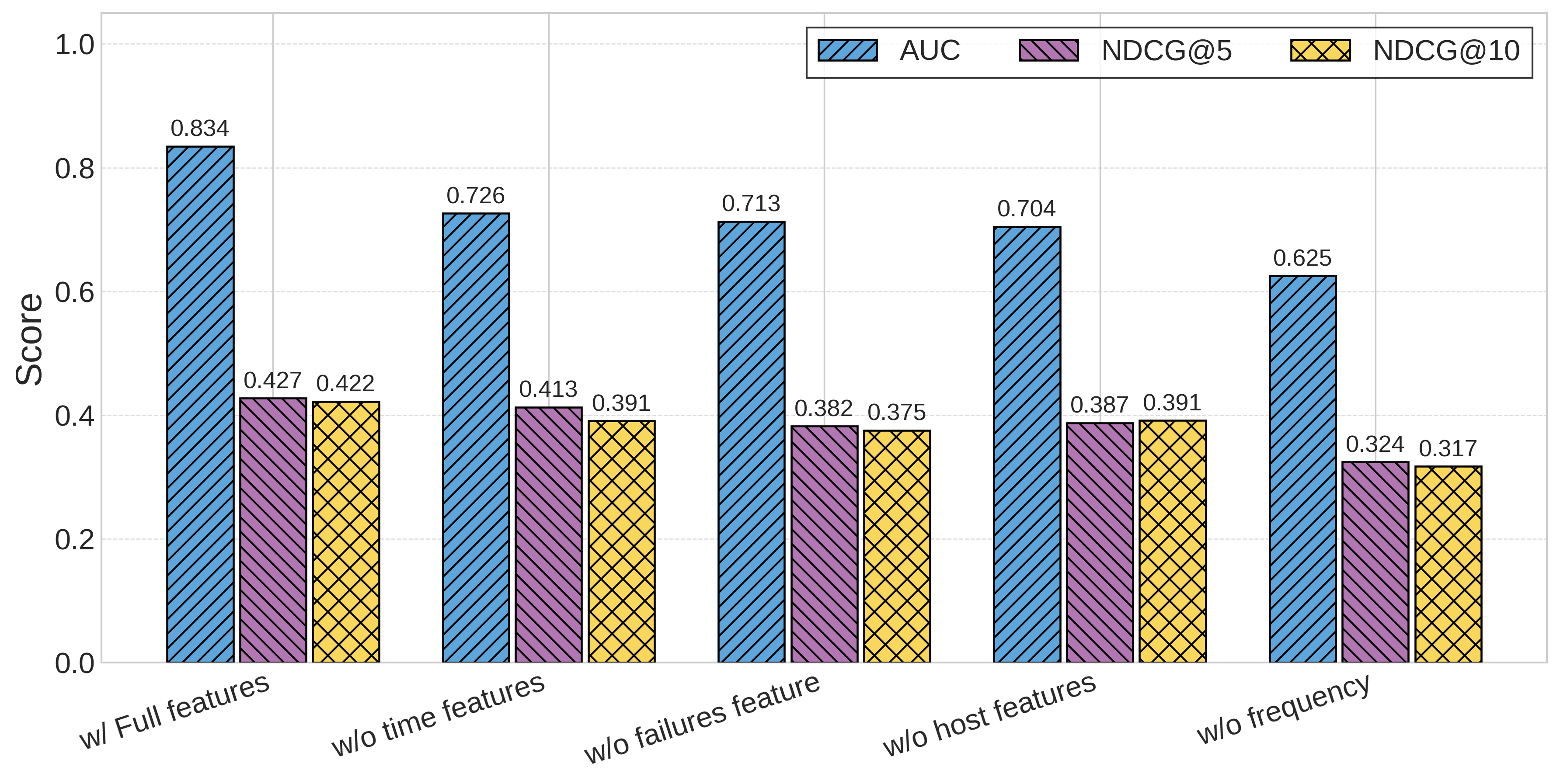}
\caption{\enspace Comparison of model performance after removing different feature types, with colors indicating different metrics.}
\vspace{-0.5cm}
\label{fig:feature importance}
\end{figure}

\noindent\textbf{Results.} Figure~\ref{fig:feature importance} summarizes the ablation trends. Overall, the most
critical signals come from \textit{recurrence and recency}: removing this group causes the largest performance drop. This indicates that short-term recurrence and history statistics (e.g., recent
fault counts, intervals, unresolved counts, and time-since-last-fault style signals) are the most informative cues for both discrimination and top-$k$ risk ranking, which aligns with the practical nature of health prediction: machines that have failed recently and repeatedly are substantially more likely to be high-risk in
the near future.

Host metadata provides the next most consistent benefit. It captures persistent susceptibility differences
across hardware/host cohorts and improves overall separability across the fleet, but is less decisive than
recency/recurrence for ordering the very top risky machines.

By comparison, failure-composition and calendar-time features contribute more modest gains. They provide
useful context (e.g., distinguishing different failure mechanisms or weak seasonality effects), yet their
incremental value is secondary once the model already observes strong recurrence and host-context signals.
Overall, the ablation study suggests that robust operational risk ranking is driven primarily by
history-derived recurrence patterns, complemented by persistent host context.

\subsection{RQ3: Sensitivity to Labeling Horizon}
Unlike standard supervised learning, GPU failure prediction has no direct ground truth for the latent ``failure probability'' $P(\textit{failure})$. In production, we only observe discrete, stochastic failure events. Consequently, constructing binary labels $y\in\{0,1\}$ using a time horizon $\Delta t$ is inevitably a human-defined heuristic that approximates this latent risk. In this experiment, we investigate how the choice of $\Delta t$ changes the nature of the learning task. Table~\ref{tab:window_comparison} shows a clear trend: ranking metrics (AUC and NDCG) improve substantially as the prediction horizon increases from 3 days to 30 days.  This trend underscores a key distinction between predicting and ranking: short horizons are dominated by stochastic triggers (aleatoric uncertainty), making precise localization infeasible. In contrast, longer horizons smooth out transient noise, allowing labels to better proxy intrinsic hardware fragility (epistemic uncertainty). Consequently, HeaRank functions less as a near-term forecaster and more as a \textit{denoising mechanism} that exposes persistent degradation.

Although a 30-day window achieves the highest AUC (0.881), it introduces a practical tension: as $\Delta t\to\infty$, risk differentiation becomes less meaningful if all nodes eventually fail. In the practical 30-day regime, the strong ranking performance indicates that HeaRank successfully sorts machines by their rate of degradation. For deployment, we select a 7-day window (AUC 0.834) not because it is the theoretical maximum, but because it provides a stable, denoised risk signal while remaining actionable for weekly maintenance scheduling.

\begin{table}[tbp]
\large
\centering
\caption{Performance comparison under different labeling horizons $\Delta t$.}
\label{tab:window_comparison}
\resizebox{\linewidth}{!}{
\begin{tabular}{lcccc}
\toprule
\textbf{Prediction Window} & \textbf{AUC} & \textbf{NDCG@5} & \textbf{NDCG@10} & \textbf{NDCG@20} \\
\midrule
3 days  & 0.723 & 0.1263 $\pm$ 0.18 & 0.1463 $\pm$ 0.19 & 0.1100 $\pm$ 0.20 \\
7 days  & 0.834 & 0.4267 $\pm$ 0.36 & 0.4225 $\pm$ 0.30 & 0.3743 $\pm$ 0.23 \\
30 days & \textbf{0.881} & \textbf{0.7219 $\pm$ 0.27} & \textbf{0.7090 $\pm$ 0.28} & \textbf{0.7504 $\pm$ 0.27} \\
\bottomrule
\end{tabular}
}
\vspace{-0.3cm}
\end{table}

\subsection{RQ4: Hybrid Telemetry and History}

A natural follow-up question is whether telemetry can complement HeaRank's history-only features. To test this, we built a hybrid ranker. For each $(\text{host}, \text{query time})$, we extracted the preceding 3 days of telemetry across 13 host-level GPU metrics, computed mean/std/min/max aggregates per metric, and concatenated these statistics with HeaRank's original failure-history features. The same pointwise MLP backbone was trained on the combined feature set. Because telemetry coverage is incomplete, we restrict the comparison to the subset of hosts with usable telemetry at each query time, and re-evaluate the history-only model on the same subset for a fair comparison.

\begin{table}[!htbp]
\centering
\small
\caption{Hybrid vs.\ history-only on the telemetry-available subset.}
\label{tab:hybrid}
\resizebox{\linewidth}{!}{
\begin{tabular}{lccc}
\toprule
\textbf{Model} & \textbf{AUC} & \textbf{NDCG@5} & \textbf{NDCG@10} \\
\midrule
History-only                              & \textbf{0.806} & 0.432 & \textbf{0.428} \\
Hybrid (history + 3-day telemetry stats)  & 0.783 & \textbf{0.452} & 0.406 \\
\bottomrule
\end{tabular}
}
\end{table}

The hybrid configuration shifts the ordering but does not produce a consistent improvement: it gains slightly at NDCG@5 but loses on AUC and NDCG@10. This is consistent with our predictability analysis in Section~\ref{sec:data_analysis}: telemetry metrics carry low SNR (Figure~\ref{fig:snr}), pre-failure precursors lack temporal localization (Figure~\ref{fig:f1trend}), and pre-failure distributions overlap with normal operation (Figure~\ref{fig:Distribution comparison}). Concatenating noisy, fine-grained telemetry aggregates with stable history features can therefore introduce additional noise on top of any signal. We view hybrid modeling as a promising but non-trivial extension that warrants more careful design, rather than something that naive feature concatenation can solve.

\section{Online Ranking Performance}
\begin{figure}[tbp]
\centering
\includegraphics[width = \linewidth]{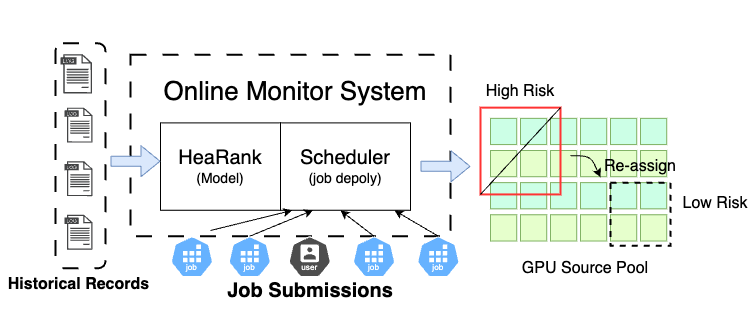}
\caption{Demonstrates how job scheduling can be done via HeaRank. When a job is submitted, the cluster's resource scheduler schedules the job assignment to a low-risk GPU server based on the ordering obtained by HeaRank.}
\label{fig:pipeline}
\vspace{-0.5cm}
\end{figure}

Figure~\ref{fig:pipeline} illustrates HeaRank's integration into the production job scheduling pipeline. The cluster scheduler assigns workloads to low-risk GPUs based on daily risk rankings, avoiding unstable nodes before failures manifest.

\begin{figure}[htbp]
\centering
\includegraphics[width= 0.8\linewidth]{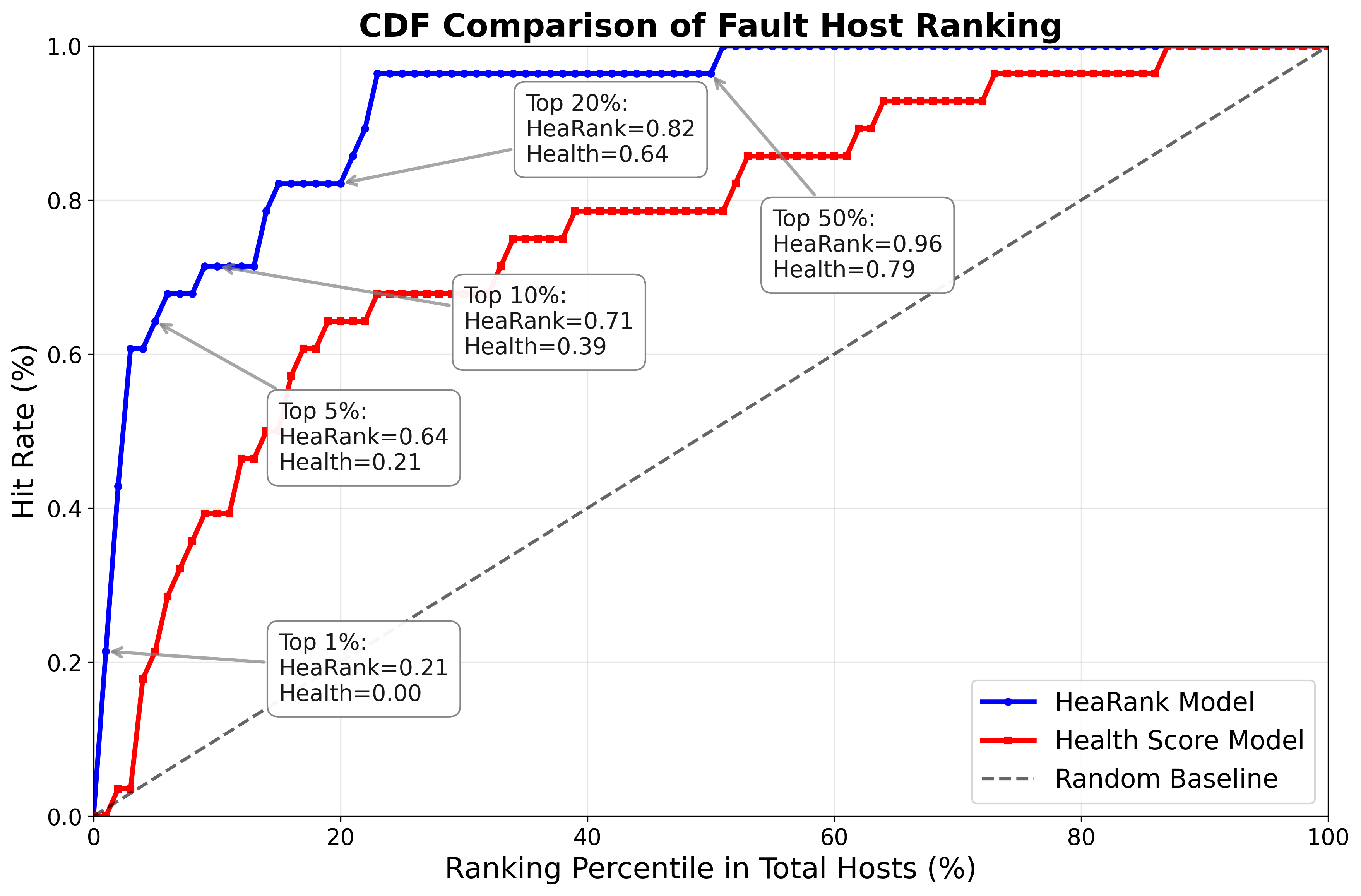}
\caption{\enspace Real-World Failure Distribution: HeaRank vs. Incumbent Health Scoring System.}
\label{fig:cdf}
\vspace{-0.4cm}
\end{figure}

We integrated HeaRank into the production scheduling pipeline of a cluster with thousands of GPUs. Unlike short-term pilots, we conducted a comprehensive six-month longitudinal study (from 2025/07 to 2026/01) to evaluate the system's stability and practical value under diverse operational conditions. Each day, HeaRank generated a complete risk ordering using historical failures, frequency patterns, and host context. Figure~\ref{fig:cdf} summarizes the aggregated results over the six-month period. Consistent with our offline experiments, failures remained highly concentrated: 64\% of all breakdown incidents observed over the half-year occurred within the top 5\% of nodes ranked by HeaRank. This concentration enables operators to focus monitoring and 
preventive maintenance on a small subset of machines while retaining broad coverage.

The operational impact is substantial. To quantify it, we provide a transparent GPU-hours back-of-the-envelope estimate rather than a single headline number, so that operators can re-derive the savings for their own fleets. Let $N_f$ denote the number of critical failures per month observed in the cluster, and let $\Delta r = 0.64 - 0.21 = 0.43$ denote HeaRank's incremental capture rate over the incumbent health-score system at the top-5\% risk tier. For a synchronous training job running on $1000$ GPUs, the expected wasted compute per failure can be written as
\begin{equation*}
\text{Loss}_{\text{per fault}} = 1000 \times \left(\frac{T_{\text{ckpt}}}{2} + T_{\text{restart}}\right) \;\;\text{GPU-hours},
\end{equation*}
where $T_{\text{ckpt}}$ is the checkpoint interval (on average half of it is lost between the last checkpoint and the failure) and $T_{\text{restart}}$ is the restart overhead. The monthly avoided GPU-hours are therefore
\begin{equation*}
\Delta \text{GPU-hours} = N_f \times \Delta r \times 1000 \times \left(\frac{T_{\text{ckpt}}}{2} + T_{\text{restart}}\right).
\end{equation*}
Using the public GPU price assumption from the DeepSeek-V3 technical report (\$2 per GPU-hour)~\citep{liu2024deepseek}, this directly translates into dollar savings that scale linearly with cluster size, failure frequency, and checkpoint cadence. Plugging in the failure frequency and checkpoint cadence observed in our cluster, this back-of-the-envelope calculation yields an estimated saving of roughly \$50{,}000 per month at our current operating scale. These results highlight both the predictive precision and the economic value of deploying HeaRank in production.

\section{Discussion}
 The success of HeaRank stems from leveraging historical failure records as a stable information source. The risk-ranked outputs allow maintenance teams to proactively investigate top-ranked nodes, often referred to as ``lemon nodes.'' They also enable schedulers to steer critical workloads away from fragile hardware. Moreover, persistent high-risk scores provide data-driven grounds for hardware retirement decisions. Beyond this core methodology, several aspects of practical deployment warrant further discussion.
\vspace{-0.2cm}
\paragraph{Cold-start and New Hardware.} 
In production environments, new hardware is typically treated with caution due to the ``infant mortality'' phenomenon, where manufacturing defects manifest early in the lifecycle. Consequently, operational protocols often default new nodes to a high-risk or probationary status rather than assuming they are healthy. HeaRank aligns with this robust strategy by incorporating static hardware metadata (e.g., GPU models, generations, and batch IDs) alongside historical failure count. This allows the model to utilize \textit{cohort analysis}: even a fresh node with zero operational history inherits a baseline risk probability, derived from its hardware group properties. Once the first real failures are observed, the history features quickly take over and the host's ranking is corrected upward; concrete examples of this two-stage behavior are provided in Appendix~\ref{appendix:cold_start}.
 
\vspace{-0.2cm}
\paragraph{Model Evolution and Online Learning.}
Hardware reliability is not a static property but a dynamic process that evolves over time. Failure modes shift as components age, and new patterns emerge as different machine types are added to the fleet. Consequently, a static model may eventually face concept drift. Regular model refreshes allow the system to adapt to the changing lifecycle of the hardware and capture the distinct failure signatures of newly introduced architectures.
\vspace{-0.2cm}
\paragraph{Orthogonality of Paradigm and new Features.} 
It is important to clarify that our proposed paradigm shift is \textit{orthogonal} to the discovery of historical feature stability.
It does not preclude the use of telemetry; rather, it provides a more robust supervision signal (relative risk) that could potentially stabilize volatile telemetry features in future hybrid models. Our contribution lies in demonstrating that even with minimal, stable historical signals, the ranking paradigm significantly outperforms traditional classification approaches.

\section{Conclusion}
This paper investigates GPU reliability in large-scale clusters and finds that accurate prediction of critical failures based solely on telemetry time-series data is inherently difficult. This is because of the characteristics of GPUs including workload-induced pattern instability, pervasive signal noise, and the lack of distinguishable signatures between failure and non-failure states. In response, we propose a shift from temporal prediction to a ranking-based approach that leverages stable historical failure information rather than alongside time-series telemetry features. Our \HeaRank model validates this approach, achieving 0.834 AUC and capturing over 64\% of actual failures within the top 5\% risk tier in production deployment. These findings suggest that future research in cluster system reliability may benefit from exploring hybrid approaches that combine multiple information sources rather than relying solely on telemetry data, particularly in environments where failure precursors exhibit high variability and randomness. 
\begin{acks}
This work was supported by the Chinese Academy of Sciences Youth Talent Program, the National Natural Science Foundation of China-Research Grants Council (RGC) Joint Research Scheme (62321166652).
\end{acks}
\bibliographystyle{ACM-Reference-Format}
\bibliography{sample-base}

\appendix
\section{System Architecture and Data Collection}

\subsection{Cluster Overview}
\begin{figure}[htbp]
\centering
\includegraphics[width = 0.95\linewidth]{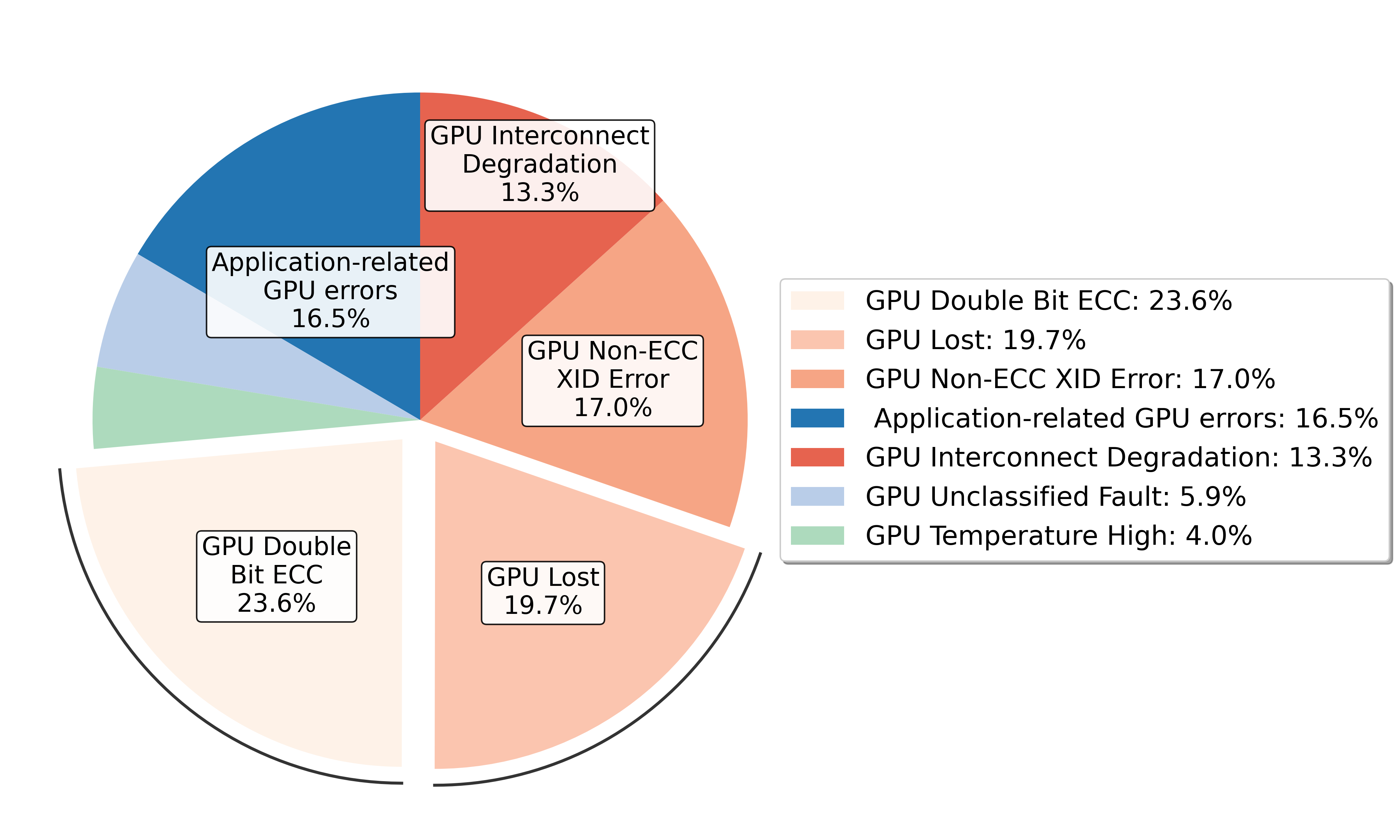}
\caption{\enspace Distribution of GPU fault types observed in our production cluster.}
\label{fig:gpu_fault_count}
\vspace{-0.4cm}
\end{figure}

Our analysis is based on a large production GPU cluster spanning thousands of GPUs across multiple hardware generations (primarily NVIDIA), designed for large-scale distributed large language model (LLM) training. Each server typically hosts eight GPUs and is interconnected via PCIe or NVLink~\citep{pcieNvidia,nvlinkNvswitch}, while jobs are orchestrated using Kubernetes. The fault statistics are collected from nearly 5,000 failures between January 2024 and December 2025. Figure~\ref{fig:gpu_fault_count} summarizes the distribution of GPU fault types observed in our production environment. The concentration on a small number of high-impact categories (e.g., DBEs and GPU Lost) motivates our focus on these failures throughout the paper.

\subsection{Telemetry Collection Pipeline}
\begin{figure}[!htbp]
\centering
\includegraphics[width = 0.75\columnwidth]{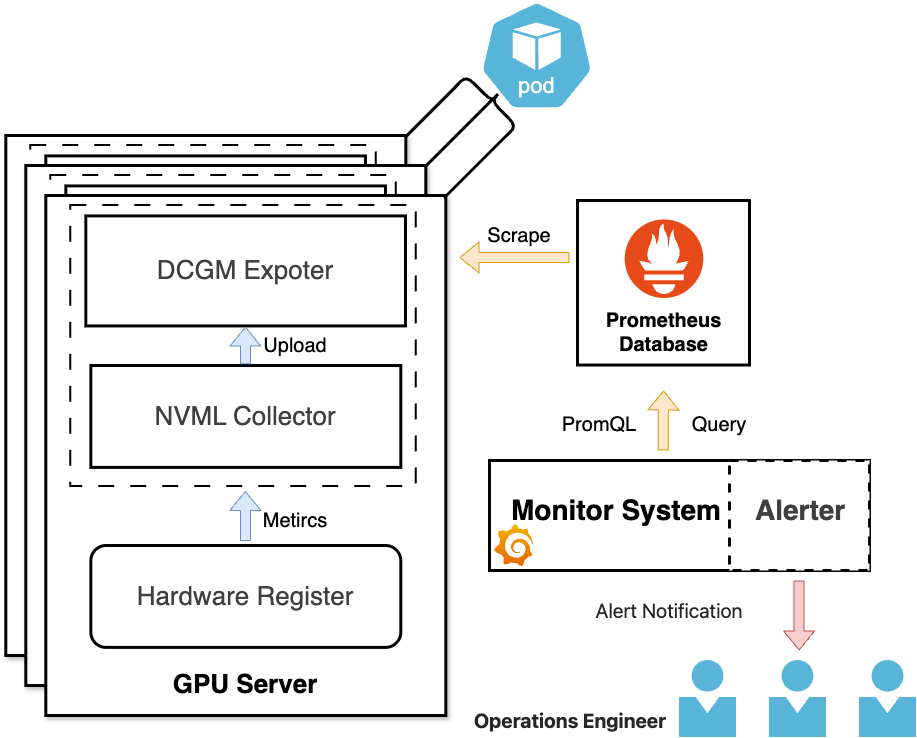}
\caption{Telemetry collection pipeline.}
\label{fig:dcgm}
\end{figure}

Figure~\ref{fig:dcgm} summarizes our observability pipeline. Each server runs a DCGM exporter that collects GPU metrics via NVML/DCGM and exposes them to Prometheus for time-series storage. Operators can inspect these metrics through Grafana, and we retrieve historical traces for specific hosts using PromQL when failures occur.

\subsection{Telemetry Dataset (for Predictability Analysis)}

To rigorously test the capabilities of conventional time-series models (e.g., LSTM, Transformer), we focused on the ``best-case'' scenario: detecting patterns for the two most dominant failure types---Double Bit Errors (DBEs) and GPU Lost events, which collectively account for 43.3\% of critical incidents.
\begin{table}[htbp]
    \centering
    \small
    \caption{Top-K F1 across prediction windows.}
    \label{tab:model_performance}
    \resizebox{\linewidth}{!}{
    \begin{tabular}{c c c c c c}
        \toprule
        \textbf{Prediction Window} & \textbf{XGBoost} & \textbf{CNN} & \textbf{LSTM} & \textbf{Transformer} & \textbf{MoE} \\ 
        \midrule
        1 hour  & 0.006  & 0.0932 & 0.0345 & 0.0562 & 0.1142 \\
        2 hour  & 0.0334 & 0.1254 & 0.0601 & 0.1026 & 0.2253 \\
        3 hour  & 0.0631 & 0.1780 & 0.1381 & 0.2254 & 0.2854 \\
        4 hour  & 0.1697 & 0.2339 & 0.1715 & 0.2035 & 0.2976 \\
        5 hour  & 0.1711 & 0.2766 & 0.2016 & 0.3012 & 0.2338 \\
        6 hour  & 0.1935 & 0.3496 & 0.3175 & 0.3149 & 0.3074 \\
        7 hour  & 0.2086 & 0.4153 & 0.3818 & 0.3768 & 0.3981 \\
        8 hour  & 0.2026 & \underline{0.4759} & 0.4190 & 0.4551 & \textbf{0.4837} \\
        \bottomrule
    \end{tabular}
    }
\end{table}
\subsubsection{Hierarchical Sampling and Balancing}
Given the extreme rarity of failures in production, using raw logs would result in a highly imbalanced dataset (positive ratio $< 0.01\%$). To ensure model sensitivity, we employed a hierarchical negative sampling strategy:
\begin{itemize}
    \item \textbf{Host-Level Balancing:} We first identified all unique machine instances ($N_{fault}$) that experienced critical failures. We then randomly sampled an equal number of healthy machines ($N_{healthy} \approx N_{fault}$) from the same clusters and hardware generations to form a balanced host set. This resulted in a dataset covering \textbf{thousands of unique GPU instances}.
    \item \textbf{Window-Level Generation:} For each selected host, we extracted high-frequency telemetry (15-second granularity). We employed a sliding window approach with a stride of 5 minutes to generate fixed-length observation sequences.
\end{itemize}

\subsubsection{Labeling and Sample Statistics}
Despite the 1:1 host balance, failure is a transient event. A positive label ($y=1$) is assigned only if the window falls within the prediction horizon (e.g., 6 hours) preceding a failure. All other windows from healthy hosts, as well as windows from faulty hosts outside the critical horizon, are labeled as negative ($y=0$).
This rigorous temporal labeling resulted in a final dataset comprising \textbf{tens of thousands of observation samples}, with a positive sample ratio of approximately \textbf{2\%}. This ratio is realistic for early warning scenarios while providing sufficient signal density for classifiers to learn if deterministic precursors exist.

\subsubsection{Preprocessing and Tuning}
The raw telemetry data underwent a standard preprocessing pipeline: timestamp alignment to true failure onset (via DCGM ECC metrics), linear interpolation for minor missing values ($< 1\%$ of samples), and Gaussian smoothing to suppress bursty sensor noise.
Critically, to address the remaining 2\% imbalance, we did not rely on default decision thresholds (0.5). Instead, we performed \textbf{threshold moving} on the validation set to maximize the F1-score for each baseline. We observed that the model output probabilities for both positive and negative samples were highly clustered (often overlapping in the 0.4--0.6 range or near zero), indicating that the models could not establish a confident decision boundary due to the low signal-to-noise ratio of the telemetry data.

\subsection{HeaRank Dataset (for Risk Ranking)
}
Based on the findings from the predictability analysis, specifically that telemetry is dominated by workload noise and lacks consistent precursors, we constructed a separate dataset for the HeaRank model that explicitly excludes time-series telemetry metrics. This dataset spans the same period (January 2024 to July 2025) but strictly utilizes historical failure logs and static metadata (as detailed in Table ~\ref{tab:features}). Unlike the short-term horizons used in the predictability analysis, this dataset employs a rolling query window. For every monitored host, we generate a query every 3 days. Each query is labeled positive ($y=1$) if the host experiences any critical failure type within the subsequent prediction horizon (e.g., $\Delta t = 7$ days), and negative ($y=0$) otherwise. This setup aligns with the operational goal of identifying "lemon nodes" based on their long-term health trajectory rather than predicting immediate signal spikes.

\begin{table}[!htbp]
    \centering
    \footnotesize
    \caption{Feature summary.}
    \label{tab:features}
    \resizebox{\linewidth}{!}{
    \begin{tabular}{llll}
        \toprule
        \textbf{Feature Category} & \textbf{Feature Name} & \textbf{Description} & \textbf{Unit} \\
        \midrule
        \multirow{13}{*}{\textbf{Time-Series Metrics}}
        & GPU Utilization & Percentage of GPU compute resource in use & \% \\
        & Frame Buffer Used & Frame buffer memory consumption & Bytes \\
        & FP16 / FP32 Activity & Ratio of half-/single-precision floating point usage & -- \\
        & SM Clock & Streaming multiprocessor clock frequency & Hz \\
        & SM Occupancy & Active warps running on SMs & \% \\
        & SM Activity & Percentage of SMs doing useful work & \% \\
        & Tensor Activity & Utilization of tensor cores & \% \\
        & Mem Copy Utilization & GPU host memory copy resource usage & \% \\
        & PCIe Tx Bytes & Bytes sent from GPU via PCIe & Bytes \\
        & PCIe Rx Bytes & Bytes received via PCIe & Bytes \\
        & GPU Temperature & Temperature sensor reading & $^\circ$C \\
        & Power Usage & Instantaneous power draw & Watts \\
        \midrule
        \multirow{9}{*}{\textbf{Host Context Features}}
        & Host Type / ID & Machine model and unique ID & Encoded \\
        & Data Center / Rack Location & Physical deployment location & Encoded \\
        & Time Context & \texttt{is\_weekend}, \texttt{day\_of\_week}, etc. & Categorical \\
        & Failure Count & Historical Failure Count (per class/type) & Count \\
        & Recent failure Stats & failures in recent window (e.g., 30 days) & Count \\
        & failure Frequency & Mean failure interval, occurrence rate & -- \\
        & Recent-to-Total Ratio & Proportion of recent to lifetime failures & -- \\
        & failure Type Breakdown & Counts by level/class/subclass & Count \\
        \bottomrule
    \end{tabular}
    }
\end{table}

\subsection{Comparison of LTR Objectives}\label{appendix:ltr_objective}

We compared three training losses under an identical configuration (same data, temporal split, and MLP backbone): pointwise sigmoid cross-entropy, listwise ListNet, and pairwise LambdaRank-style.

\begin{table}[!htbp]
\centering
\small
\caption{Effect of the training objective under the same backbone and data split.}
\label{tab:ltr_objective}
\resizebox{\linewidth}{!}{
\begin{tabular}{lcccc}
\toprule
\textbf{Loss} & \textbf{AUC} & \textbf{NDCG@5} & \textbf{NDCG@10} & \textbf{NDCG@20} \\
\midrule
Pointwise (BCE)              & \textbf{0.834} & \textbf{0.4398} & \textbf{0.4185} & \textbf{0.3818} \\
Listwise (ListNet)           & 0.812 & 0.3559 & 0.3444 & 0.3023 \\
Pairwise (LambdaRank-style)  & 0.779 & 0.2656 & 0.2716 & 0.2578 \\
\bottomrule
\end{tabular}
}
\end{table}

Pointwise wins on AUC and all top-$K$ cutoffs. Two practical reasons: (i) our supervision is absolute (per-host binary labels), not relative pairwise rankings, so pointwise BCE consumes the signal directly without the noise introduced by pair/group construction; (ii) pointwise scoring preserves a calibrated probability interpretation, which is required for stable threshold-based operational policies.

\subsection{Cold-Start Examples}\label{appendix:cold_start}

To illustrate how HeaRank handles hosts without any historical failure record, we tracked hosts in the test window that experience their \emph{first} failure during evaluation. When \texttt{history\_fault\_count} and \texttt{recent\_fault\_count} are both zero, the historical features carry no positive signal; the score in this regime is driven entirely by static metadata (\texttt{host\_model}, \texttt{gpu\_model}, \texttt{manufacturer}, \texttt{main\_board}, \texttt{quota\_group}), i.e., a \emph{cohort-level prior risk}. Table~\ref{tab:cold_start} shows two representative cases. In both, the host is ranked clearly above the cluster median even before any failure has been recorded (e.g., Host2 sits at 18.5\%); once the first real failure is observed, history features take over and the rank is rapidly pulled to the top (1.0\% / 1.9\%).

\begin{table}[!htbp]
\centering
\small
\caption{Cold-start ranking behavior of HeaRank before vs.\ after the first observed failure.}
\label{tab:cold_start}
\begin{tabular}{llcc}
\toprule
\textbf{Host} & \textbf{Status} & \textbf{Score} & \textbf{Rank (\%)} \\
\midrule
Host1 & no history, label=1     & $-2.40$ & 50.8\% \\
Host1 & first failure observed  & $-0.58$ & 1.0\% \\
Host2 & no history, label=1     & $-1.72$ & 18.5\% \\
Host2 & first failure observed  & $-0.62$ & 1.9\% \\
\bottomrule
\end{tabular}
\end{table}

These examples show that HeaRank does not naively treat history-less hosts as low risk: cohort metadata already pulls genuinely fragile new hosts above the median, and a single observed failure is enough to push them into the top-$K$ tier.
\subsection{Open-Source Release}
The code for this project has been open-sourced to facilitate reproducibility and further research. Due to enterprise privacy and compliance policies, we are unable to release the production dataset used in this study. We provide a failure record template that specifies the required fields and aggregation format. Any dataset collected and summarized following this template can be used to train and test our model.\noindent\textbf{Repository:} \texttt{\url{https://github.com/geotle77/kdd26-artifact}}
\end{document}